\documentclass[preprint,prd,superscriptaddress,nofootinbib,floatfix]{revtex4}
\usepackage[dvipdfmx]{graphicx}
\usepackage[dvipdfmx]{color}
\usepackage{bm}
\usepackage{multirow}

\begin{document}

\preprint{OU-HET-869, KEK-CP-324, YITP-15-69}

\title{
$\eta^\prime$ meson mass from topological charge density correlator in QCD 
}

\newcommand{\Tsukuba}{
  Graduate School of Pure and Applied Sciences,
University of Tsukuba, Tsukuba, Ibaraki 305-8571, Japan
}

\newcommand{\CCS}{
Center for Computational Sciences,
University of Tsukuba, Ibaraki 305-8577, Japan
}
\newcommand{\KEK}{
  KEK Theory Center,
  High Energy Accelerator Research Organization (KEK),
  Tsukuba 305-0801, Japan
}
\newcommand{\GUAS}{
  School of High Energy Accelerator Science,
  The Graduate University for Advanced Studies (Sokendai),
  Tsukuba 305-0801, Japan
}
\newcommand{\YITP}{
  Yukawa Institute for Theoretical Physics, 
  Kyoto University, Kyoto 606-8502, Japan
}
\newcommand{\Osaka}{
  Department of Physics, Osaka University,
  Toyonaka 560-0043, Japan
}

\author{H.~Fukaya}
\affiliation{\Osaka}

\author{S.~Aoki}
\affiliation{\YITP}
\affiliation{\CCS}

\author{G.~Cossu}
\affiliation{\KEK}

\author{S.~Hashimoto}
\affiliation{\KEK}
\affiliation{\GUAS}

\author{T.~Kaneko}
\affiliation{\KEK}
\affiliation{\GUAS}

\author{J.~Noaki}
\affiliation{\KEK}


\collaboration{JLQCD collaboration}
\noaffiliation

\begin{abstract}
The flavor-singlet component of the $\eta^\prime$ meson is related to the
topological structure of the SU(3) gauge field through the chiral anomaly.
We perform a $2+1$-flavor lattice QCD calculation and
demonstrate that the two-point function of
a gluonically defined topological charge density 
after a short Yang-Mills gradient flow
contains the propagation of the $\eta^\prime$ meson,
by showing that its mass in the chiral and continuum limit
is consistent with the experimental value.
The gluonic correlator does not suffer from the contamination
of the pion contribution, and the clean signal is obtained
at significantly lower numerical cost compared to the conventional
method with the quark bilinear operators.
\end{abstract}

\maketitle

Among other hadrons the $\eta^\prime$ meson plays a special role in
the study of the topological structure of Quantum Chromodynamics (QCD).
The $\eta^\prime$ meson would be a pseudo Nambu Goldstone boson
associated with the spontaneous breaking of the axial $U(1)$ symmetry,
while it acquires a large mass through the chiral anomaly \cite{Weinberg:1975ui}, 
which relates the divergence of the flavor-singlet axial vector current
to the topological charge density 
in QCD.
Witten \cite{Witten:1979vv} and Veneziano \cite{Veneziano:1980xs}
estimated the $\eta^\prime$ meson mass in the large-$N_c$ 
(number of colors) limit and showed that 
its mass squared is proportional to 
the topological susceptibility of QCD.

In real QCD with $N_c=3$ and light dynamical quarks,
the argument of Witten and Veneziano is no longer valid.
It is not the $\eta^\prime$ meson but the pion 
that governs the topological susceptibility.
In fact, in our previous lattice QCD simulations where
we kept the chiral symmetry (nearly) exact
\cite{Aoki:2007pw, Hsieh:2009zz, Fukaya:2014zda},
it was confirmed that the topological susceptibility 
is proportional to the light sea quark masses as predicted by chiral perturbation theory,
$\chi_t = \frac{\Sigma}{\sum_i 1/m_i},$
where $\Sigma$ denotes the chiral condensate, and $m_i$ denotes
the $i$-th light quark mass. 
In particular, $\chi_t$ vanishes in the limit of massless up and down quarks,
reflecting 
the long-range dynamics of the pion field.

Then, an interesting question arises: what happens to the $\eta^\prime$ meson with $N_c=3$?
Since the effect of the anomaly is stronger than in the large-$N_c$ limit,
the $\eta^\prime$ meson mass should be still generated by the
topological fluctuation of the gluons.
Nevertheless, it must be insensitive to $\chi_t$.
This implies a non-trivial double-scale structure in 
the topological excitation of gauge field:
it creates the $\eta^\prime$ meson at short distances,
while making a tight connection to the pion at long distances.
The answer to this question could be that the topological property of 
the $\eta^\prime$ meson in QCD is hidden inside the pion clouds.

In this work, by an explicit calculation in $2+1$-flavor lattice QCD, 
we show that the two-point function of the topological charge density at short distances
gives a mass consistent with the experimental value of the $\eta^\prime$ meson mass.
Since we have computed $\chi_t$ using the same correlation functions
(see \cite{Fukaya:2014zda} for the details),
our data clearly show the double-scale structure
of the topological property of gluons: the long-range physics
described by the pion, and the short-range (or first excited) physics
governed by the $\eta^\prime$ meson.

Not only is it theoretically interesting, but our work also 
proposes a practically advantageous method to determine the $\eta^\prime$ meson mass.
Direct lattice calculation of the $\eta^\prime$ meson mass has been
challenging because one has to include the disconnected quark-line
diagram, which appears from the Wick contraction of the fermion
bilinear operators representing the $J^{PC}=0^{-+}$ flavor-singlet
state, as has been done in recent calculations
\cite{Kaneko:2009za,Christ:2010dd,Gregory:2011sg,Michael:2013gka,Ottnad:2015hva}.
This is numerically demanding and
statistically very noisy.

The advantage of the gluonic calculation adopted in this work over the
conventional fermionic one is two-fold.
First, we can avoid the enormous computational cost of 
stochastically evaluating the disconnected diagram.
The gluonic definition of the topological charge density does not
require any inversion of the Dirac operator.

Second, our method avoids the contamination from the pions.
In the conventional method where one calculates quark connected and
disconnected diagram contributions appearing from the Wick contraction
of quark fields, the pion may propagate in each contribution 
but it cancels between them.
A large statistics is required for a sufficient cancellation 
before extracting the $\eta^\prime$ meson propagator.
Since the purely gluonic definition of the topological charge density operator
\begin{equation}
  \label{eq:q}
  q(x) = \frac{1}{32\pi^2}\epsilon_{\mu\nu\rho\sigma}
  {\rm Tr}F_{\rm cl}^{\mu\nu}F_{\rm cl}^{\rho\sigma}(x),
\end{equation}
where $F_{\rm cl}^{\mu\nu}$ denotes the field strength tensor of
the gluon field defined through the so-called clover term consisting
of four plaquettes, does not directly couple to the pions,
its correlator is free from the pion background.

Note here that the sum of (\ref{eq:q}) over the lattice volume
gives the global topological charge up to discretization effects,
$Q=\sum_x q(x)+{\cal O}(a^2)$. However, it is well-known
that the ${\cal O}(a^2)$ contribution is large with currently available lattice spacings.
In order to reduce this statistical noise,
we modify the link variables using the Yang-Mills (YM) gradient flow \cite{Luscher:2010iy}.
At a flow time $t$, it amounts to smearing the gauge fields in a range
of the length $\sqrt{8t}$.
It is shown that the topological charge $Q$ defined through
(\ref{eq:q}) converges to an integer value at a sufficiently large
flow time \cite{Luscher:2011bx,Bonati:2014tqa}.
This smearing procedure eliminates short-distance fluctuations and
also suppresses the noise at longer distances.\footnote{
  A similar method was applied in a quenched study
  to extract the ``pseudoscalar glueball mass''
  \cite{Chowdhury:2014mra}.
  There are other viable definitions of the smearing as used in
  previous works to probe the topological structure of the QCD vacuum 
  \cite{de Forcrand:1997sq,Horvath:2005cv,Ilgenfritz:2007xu, Alles:2007zz}.
}

For our purpose of extracting the
$\eta^\prime$ meson mass, the YM gradient flow time has to be short in order 
not to distort the correlation of the $\eta^\prime$ propagation.
Assuming a simple Gaussian form of the smearing effect,
Bruno {\it et al.} \cite{Bruno:2014ova} 
estimated the size of distortion of the correlator as
\begin{eqnarray}
\frac{\Delta \langle q(x)q(y)\rangle}{\langle q(x)q(y)\rangle} \sim e^{-(|x-y|/\sqrt{8t}-m_{\eta^\prime}\sqrt{8t})^2}\frac{m_{\eta^\prime} (8t)^{3/2}}{2\sqrt{\pi}|x-y|^2}.
\end{eqnarray}
In our analysis below, we use the reference flow 
time around $\sqrt{8t}=0.2$ fm
for the fit range $|x-y|>0.6$ fm, 
which makes the above correction less than 
$1$\% for $m_{\eta^\prime}\simeq 1$ GeV.\\

We employ the Symanzik gauge action and the M\"obius domain-wall
fermion action for gauge ensemble generations \cite{Kaneko:2013jla, Noaki:2014ura, Cossu:2013ola}.
We apply three steps of stout smearing of the gauge links before
inserting it in the Dirac operator.
Our main runs of $2+1$-flavor lattice QCD simulations are carried out 
on two different lattice volumes
$L^3\times T=32^3\times 64$ and $48^3\times 96$, for which
we set $\beta$ = 4.17 and 4.35, respectively.
The inverse lattice spacing $1/a$ is estimated to be 2.4~GeV (for $\beta=4.17$) and 3.6~GeV (for $\beta=4.35$),
using the input $\sqrt{t_0}=0.1465$ fm \cite{Borsanyi:2012zs}
where the reference YM gradient flow time $t_0$ is defined 
by $t^2\langle E\rangle |_{t=t_0}=0.3$ \cite{Luscher:2010iy},
with the energy density $E$ of the gluon field.
Our two lattices share a similar physical size $L\sim 2.6$ fm.
For the quark mass, we use two values of 
the strange quark mass $m_s$ around its physical point, 
and 3--4 values of the up and down quark mass $m_{ud}$ for each $m_s$.
The lightest pion mass is around 230 MeV with 
our smallest value of $am_{ud}$ = 0.0035 at $\beta$ = 4.17.
In order to check the systematics due to finite volume sizes 
and lattice spacings, we also perform simulations on 
a larger lattice $48^3\times 96$ (at $\beta=4.17$ and $m_\pi \sim 230$ MeV),
and a finer lattice $64^3\times 128$ (at $\beta=4.47$ [$1/a\sim 4.5$ GeV] and $m_\pi \sim$ 285 MeV).
For each ensemble, 500--1,000 gauge configurations are sampled
from 10,000 molecular dynamics (MD) time.
%
The residual mass in the M\"obius domain-wall fermion formalism is
kept smaller than $\sim$ 0.5~MeV \cite{Hashimoto:2014gta}
by choosing $L_s$ = 12 at $\beta$ = 4.17
and $L_s$ =  8  at $\beta$ = 4.35 (and 4.47).

On each generated configuration, we perform 500--1,000 steps of the
YM gradient flow (using the conventional Wilson gauge action) with a step-size $a^2\Delta t=$0.01. 
At every 20--30 steps, we store  $q(x)$ 
and measure its correlator using the Fast Fourier Transform (FFT) technique.
As reported in \cite{Fukaya:2014zda}, the flow time history
of the gluonic definition of the topological charge $Q$ 
shows a good convergence to discrete values near integers.


We find that the two-point function $\langle q(x)q(y)\rangle$ 
at our target distance $|x-y|\sim 0.7$ fm always
shows a shorter autocorrelation time than 10 MD time,
while its global average, $Q=\sum_x q(x)$ 
has $O(100)$ or higher MD time at $\beta=4.35$.
This is a good evidence that the $\eta^\prime$ meson physics
is {\it decoupled} \cite{Schaefer:2010hu} from the global topological charge. 
In the following analysis, we estimate the statistical error
by the jackknife method after binning the data in 140--200 MD time.

\begin{figure*}[tbp]
  \centering
  \includegraphics[width=10cm]{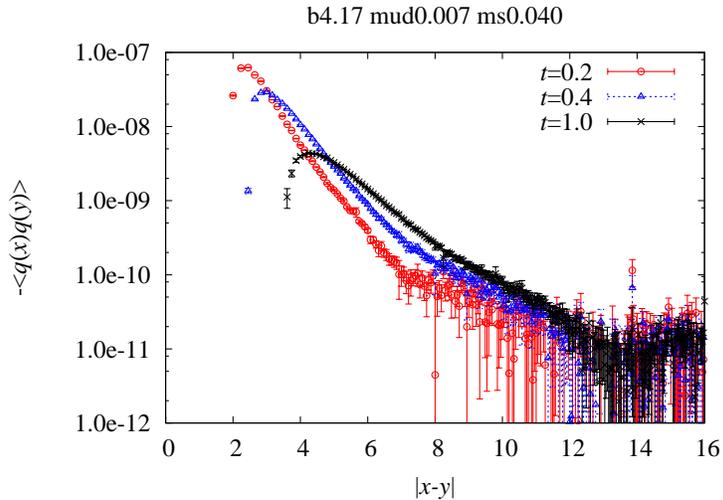}
  \caption{
    The correlator $-\langle q(x)q(y) \rangle$  at
    the flow times $a^2t$ = 0.2 (circles), 0.4 (triangles) and 
    1.0 (crosses). 
    Data at $\beta=4.17$, $am_{ud}=0.007$ and $am_s=0.040$ are presented.
  }
  \label{fig:corr}
\end{figure*}

Figure~\ref{fig:corr} shows the topological charge density correlator
$C(|x-y|)=-\langle q(x)q(y)\rangle$ at the flow times 
$a^2t$ = 0.2, 0.4 and 1.0. 
Using the FFT, data points are averaged over all possible combinations of
$x$ and $y$ giving the same $r=|x-y|$ to improve the signal.
As the flow-time increases, the statistical fluctuation of the
correlator becomes milder, while the region at small $|x-y|$ is distorted.
We therefore need to find a region of $t$ where the correlator
has sufficiently small noises to find the signal 
while it is not distorted by the smearing of the YM gradient flow.

The data for $C(|x-y|)$ show a cleaner signal than the conventional zero-momentum projection,
$\displaystyle\sum_{\vec{x}}\langle q(x)q(y)\rangle$,
because of the average in all four-dimensional directions.
The $\eta^\prime$ meson mass is extracted by fitting 
the data to the function of a single boson propagation:
\begin{eqnarray}
  f(r, m_{\eta^\prime})=A \frac{K_1(m_{\eta^\prime}r)}{r},
\end{eqnarray}
where $r=|x-y|$, 
$K_1$ is the modified Bessel function and $A$ is an unknown constant, which depends on the flow time $t$.

\begin{figure*}[tbp]
  \centering
  \includegraphics[width=10cm]{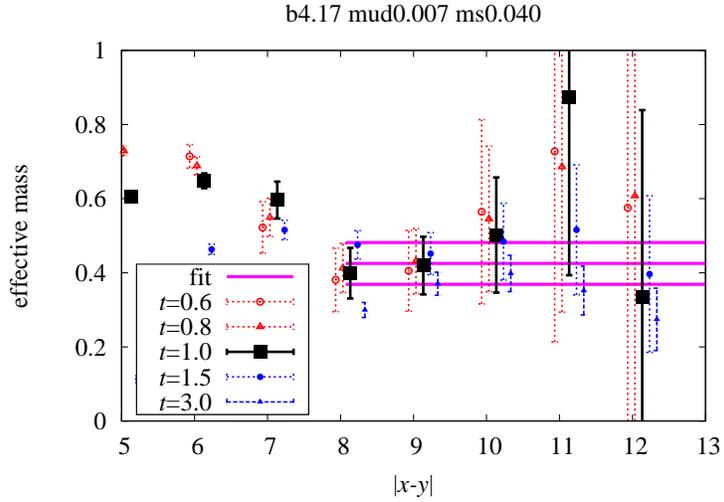}
  \caption{
    The effective mass $m_{\rm eff}(r)$ for the data with several flow
    times. 
    Results for the $\beta=4.17$  $am_{ud}=0.007$  and $am_s=0.040$ ensemble are
    shown. 
  }
  \label{fig:corr_eff}
\end{figure*}

In order to determine the fitting range,
we define a local ``effective mass'' $m_{\rm eff}(r)$
by numerically solving 
$f(r+\Delta r, m_{\rm eff}(r))/f(r, m_{\rm eff}(r))
=C(r+\Delta r)/C(r)$.
We set the interval to be $\Delta r=1/2$, 
and the data of $C(r)$ in the range $[r,r+\Delta r]$ are averaged.
As shown in Fig.~\ref{fig:corr_eff}, 
a reasonable plateau is found for $m_{\rm eff}(r)$ 
around $r\sim$ 8--12 ($> 0.6$ fm) at $t=1$ ($\sqrt{8t}\sim 0.2$ fm).

\begin{figure*}[tbp]
  \centering
  \includegraphics[width=10cm]{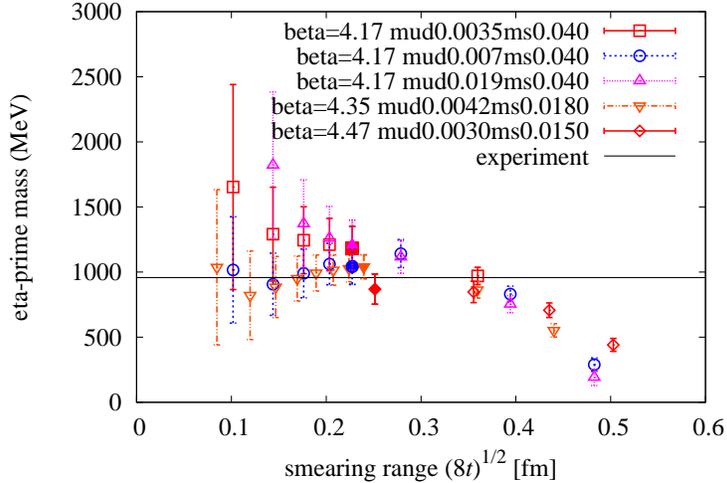}
  \caption{
    The flow-time dependence of the $\eta^\prime$ meson mass. 
    The data at various sea quark masses and $\beta$ values are shown,
    as specified in the legend. The filled symbols represent our data taken for the
    central values.
  }
  \label{fig:etapmass_vs_t}
\end{figure*}

Figure~\ref{fig:etapmass_vs_t} shows the $\eta^\prime$ meson mass 
obtained by fitting in the range [8,12] as a function of $\sqrt{8t}$.
The data around $\sqrt{8t}\sim$ 0.2~fm are stable.
At larger smearing lengths $\sqrt{8t}\gtrsim$ 0.3~fm,
we observe a large distortion of the data.
We take the data at $\sqrt{8t}=$ 0.2--0.25~fm 
(filled symbols in Fig.~\ref{fig:etapmass_vs_t}) for our results.

\begin{figure*}[tbp]
  \centering
  \includegraphics[width=10cm]{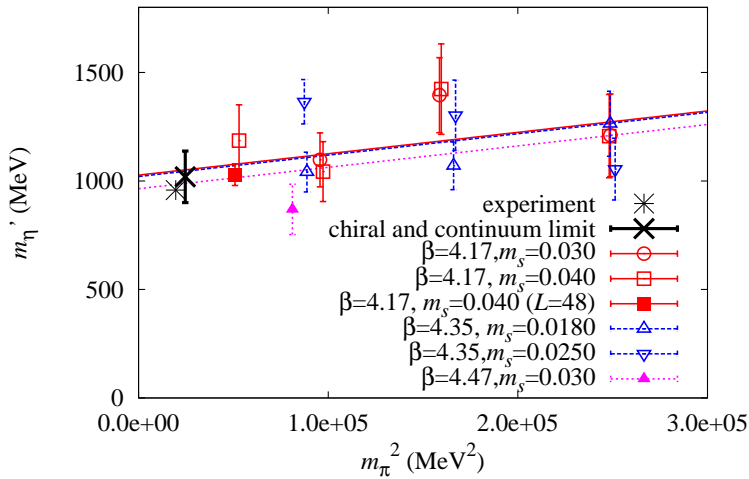}
  \caption{
    The extracted $\eta^\prime$ meson mass from each gauge ensemble.
    The three fit lines (representing the same linear fit function at three values of $a$)
    are shown for higher $m_s$'s at each $\beta$. 
}
\label{fig:etapmass}
\end{figure*}

We plot all the results in Fig.~\ref{fig:etapmass} as a function of
the square of the pion mass $m_\pi$.
We find that the dependence on the quark masses, as well as on $V$ and $a$ is mild.
We therefore perform a global fit of our data 
to a linear function 
$m_{\eta^\prime}^{\rm phys} + C_a a^2 
+ C_{ud} [m_\pi^2-(m_\pi^{\rm phys})^2] + C_s[(2m_K^2-m_\pi^2)-\{2(m_K^{\rm phys})^2-(m_\pi^{\rm phys})^2\}]$,
where $m_{\eta^\prime}^{\rm phys},\; C_a,\; C_{ud}$, and $C_s$ 
are taken as free parameters, and $m_{\pi/K}^{\rm phys }$ 
denotes the experimental value of the pion/kaon mass.
As shown by the lines (which are shown for higher $m_s$ only)
in Fig.~\ref{fig:etapmass},
we find that our linear function fits the lattice data
reasonably with $\chi^2$/(degrees of freedom) $\sim$ 1.6.

 
Because of possible bias in the topological charge sampling, the
$\eta^\prime$ correlator may not decay exponentially but become
a constant at long distances \cite{Bali:2014pva}.
In each topological sector $Q$, it is predicted as \cite{Aoki:2007ka} 
\begin{equation}
\label{eq:Q2correction}
\langle q(x)q(y)\rangle^Q \sim 
\frac{1}{V}\left(\frac{Q^2}{V}-\chi_t\right)\;\;\;\mbox{at large $|x-y|$}.
\end{equation}
The typical size of the constant term $\chi_t/V$ is 100-1000 times
smaller than $|\langle q(x)q(y)\rangle |$ in our fitting range, 
which is consistent with the fact that this observable shows
no strong correlation to the global topological charge $Q$.

We also estimate that the finite volume effect on our observable
is negligible, since the $\eta^\prime$ meson propagator 
implies $\exp(-m_{\eta^\prime}L) \sim 3\times 10^{-6} $
even at our lightest $m_{ud}$. Our data on a larger lattice $48^3\times 96$,
which are statistically consistent with those on the smaller lattice,
support this assumption.

\begin{figure*}[tbp]
  \centering
  \includegraphics[width=10cm]{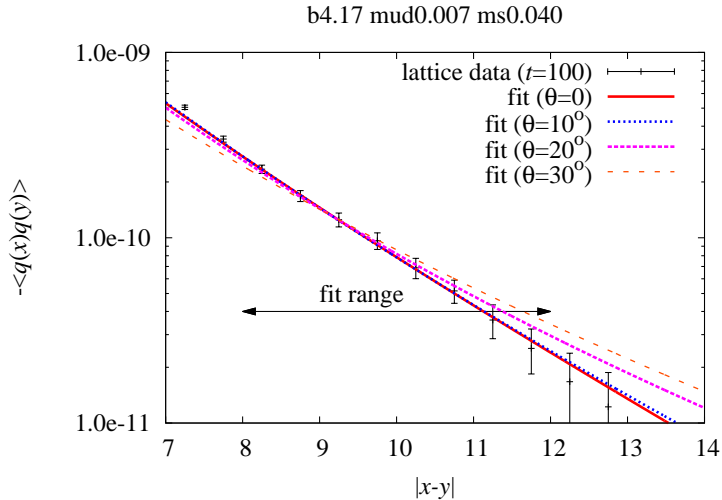}
  \caption{
    Comparison of the fit curves at different input $\theta$.
  }
\label{fig:mixing_effect}
\end{figure*}

Since $m_{ud}$ and $m_s$ are non-degenerate, the flavor-singlet
channel does not correspond to the mass eigenstate and 
a mixing with the flavor non-singlet channel is expected, which is the
$\eta$-$\eta^\prime$ mixing.
We estimate the size of the $\eta$-meson contribution to the
flavor-singlet channel by modifying the fit function as
\begin{equation}
  \label{eq:etamixing}
  \frac{A}{r}K_1(m_{\eta^\prime}r) \to 
  \frac{A}{r} \left[
  e^{2m_{\eta^\prime}^2 t}m_{\eta^\prime}K_1(m_{\eta^\prime}r) +
  e^{2m_{\eta}^2 t}m_{\eta}K_1(m_{\eta}r)\tan^2 \theta
  \right],
\end{equation}
where $\theta$ is the mixing angle,
and $m_\eta$ denotes the $\eta$ meson mass.
Note that the factors $e^{2m_{\eta^\prime}^2 t}$ and
$e^{2m_{\eta}^2 t}$ come from the effect of the YM gradient flow smearing.
As a phenomenological estimate for $m_\eta$,
we use $\sqrt{m_{\eta_S}^2(m_{ud}+2m_s)/(3 m_s)}$ where
$m_{\eta_S}$ is the (unphysical) mass of the connected pseudoscalar correlator
of strange and anti-strange quarks \cite{Brendan}.
We compare the fit curves with different fixed values of $\theta$
as shown in Fig.~\ref{fig:mixing_effect}.
As $\theta$ increases the quality of the fit becomes
worse, especially at long distances.
Thus, our data suggest a small mixing angle 
$|\theta| \lesssim 10^\circ$,
and this tendency is seen at all our simulated parameters. 
(the sign of $\theta$ is not relevant for this conclusion.)
The small mixing angle is consistent with the estimates from the
quark model, $\theta=-25^\circ$-- $-10^\circ$ \cite{Beringer:1900zz}.
In our simulations, $\theta$ is likely to be smaller than these
phenomenological values since our simulated $m_{ud}$ is closer to $m_s$.
We confirm that the fit results with the fixed value of $\theta=10^o$ are
consistent with those with $\theta=0$ within the statistical errors,
although there exists a tendency that the $m_{\eta^\prime}$ meson mass
becomes higher for $\theta\neq 0$. 
We take this $\sim +5$\% deviation
as the systematic error from the $\eta$ meson mixing.

Finally, we examine the systematics in the chiral and continuum extrapolation.
Since both of the $m_\pi$ and $a^2$ dependences are mild,
even if we totally ignore these dependences, 
the (constant) fit works well, giving a 8\% different
value of $m_{\eta^\prime}^{\rm phys}$ from the original linear fit,
which is within the statistical error. 
We take this $\pm 8$\% as the possible systematic error in the extrapolations.

Our final result at the physical point is
\begin{equation}
  m_{\eta^\prime} = 1019(119)(^{+97}_{-86})\;\;\;\mbox{MeV},
\end{equation}
which is consistent with the experimental value $m_{\eta^\prime}=957.78(6)$ MeV \cite{Beringer:1900zz}.
Here the first error is statistical and the second
is the systematic error from the mixing with the $\eta$ meson
and the chiral and continuum extrapolations 
(added in quadrature).
From the same set of correlators we obtain the topological susceptibility
$\chi_t$, which will be presented elsewhere.

\vspace*{5mm}
We thank T.~Izubuchi, P.~de Forcrand, H.~Ohki and other 
members of JLQCD collaboration for fruitful discussions.
We also thank the Yukawa Institute for Theoretical Physics, 
Kyoto University. Discussions during the YITP workshop YITP-T-14-03 
on ``Hadrons and Hadron Interactions in QCD'' were useful in completing this work.
Numerical simulations are performed on IBM System Blue Gene Solution at KEK under 
a support of its Large Scale Simulation Program (No. 14/15-10). 
This work is supported in part by the Grand-in-Aid of the Japanese Ministry of Education 
(No.25287046, 25800147, 26247043, 26400259, 15K05065), and supported in part by MEXT SPIRE and JiCFuS.


\begin{thebibliography}{99}

\bibitem{Weinberg:1975ui} 
  S.~Weinberg,
  Phys.\ Rev.\ D {\bf 11}, 3583 (1975).

\bibitem{Witten:1979vv} 
  E.~Witten,
  Nucl.\ Phys.\ B {\bf 156}, 269 (1979).

\bibitem{Veneziano:1980xs} 
  G.~Veneziano,
  Phys.\ Lett.\ B {\bf 95}, 90 (1980).



\bibitem{Aoki:2007pw} 
  S.~Aoki {\it et al.}  [JLQCD and TWQCD Collaborations],
  Phys.\ Lett.\ B {\bf 665}, 294 (2008)
  [arXiv:0710.1130 [hep-lat]].

\bibitem{Hsieh:2009zz} 
  T.~H.~Hsieh {\it et al.} [JLQCD and TWQCD Collaborations],
  PoS LAT {\bf 2009}, 085 (2009).

\bibitem{Fukaya:2014zda} 
  H.~Fukaya {\it et al.} [JLQCD Collaboration],
  PoS LATTICE {\bf 2014}, 323 (2014)
  [arXiv:1411.1473 [hep-lat]].

\bibitem{Kaneko:2009za} 
  T.~Kaneko {\it et al.}  [TWQCD and JLQCD Collaborations],
  PoS LAT {\bf 2009}, 107 (2009)
  [arXiv:0910.4648 [hep-lat]].

\bibitem{Christ:2010dd} 
  N.~H.~Christ, C.~Dawson, T.~Izubuchi, C.~Jung, Q.~Liu, R.~D.~Mawhinney, C.~T.~Sachrajda and A.~Soni {\it et al.},
  Phys.\ Rev.\ Lett.\  {\bf 105}, 241601 (2010)
  [arXiv:1002.2999 [hep-lat]].

\bibitem{Gregory:2011sg} 
  E.~B.~Gregory {\it et al.}  [UKQCD Collaboration],
  Phys.\ Rev.\ D {\bf 86}, 014504 (2012)
  [arXiv:1112.4384 [hep-lat]].

\bibitem{Michael:2013gka} 
  C.~Michael {\it et al.}  [ETM Collaboration],
  Phys.\ Rev.\ Lett.\  {\bf 111}, no. 18, 181602 (2013)
  [arXiv:1310.1207 [hep-lat]].

\bibitem{Ottnad:2015hva} 
  K.~Ottnad {\it et al.} [OTM Collaboration],
  Nucl.\ Phys.\ B {\bf 896}, 470 (2015)
  doi:10.1016/j.nuclphysb.2015.05.001
  [arXiv:1501.02645 [hep-lat]].


\bibitem{Luscher:2010iy} 
  M.~L\"uscher,
  JHEP {\bf 1008}, 071 (2010)
  [Erratum-ibid.\  {\bf 1403}, 092 (2014)]
  [arXiv:1006.4518 [hep-lat]].

\bibitem{Luscher:2011bx} 
  M.~L\"uscher and P.~Weisz,
  JHEP {\bf 1102}, 051 (2011)
  [arXiv:1101.0963 [hep-th]].


\bibitem{Bonati:2014tqa} 
  C.~Bonati and M.~D'Elia,
  Phys.\ Rev.\ D {\bf 89}, no. 10, 105005 (2014)
  [arXiv:1401.2441 [hep-lat]].



\bibitem{Bruno:2014ova} 
  M.~Bruno {\it et al.}  [ALPHA Collaboration],
  JHEP {\bf 1408}, 150 (2014)
  [arXiv:1406.5363 [hep-lat]].



\bibitem{Chowdhury:2014mra} 
  A.~Chowdhury, A.~Harindranath and J.~Maiti,
  Phys.\ Rev.\ D {\bf 91}, no. 7, 074507 (2015)
  doi:10.1103/PhysRevD.91.074507
  [arXiv:1409.6459 [hep-lat]].

\bibitem{de Forcrand:1997sq} 
  P.~de Forcrand, M.~Garcia Perez and I.~O.~Stamatescu,
  Nucl.\ Phys.\ B {\bf 499}, 409 (1997)
  [hep-lat/9701012].

\bibitem{Horvath:2005cv} 
  I.~Horvath, A.~Alexandru, J.~B.~Zhang, Y.~Chen, S.~J.~Dong, T.~Draper, K.~F.~Liu and N.~Mathur {\it et al.},
  Phys.\ Lett.\ B {\bf 617}, 49 (2005)
  [hep-lat/0504005].

\bibitem{Ilgenfritz:2007xu} 
  E.-M.~Ilgenfritz, K.~Koller, Y.~Koma, G.~Schierholz, T.~Streuer and V.~Weinberg,
  Phys.\ Rev.\ D {\bf 76}, 034506 (2007)
  [arXiv:0705.0018 [hep-lat]].

\bibitem{Alles:2007zz} 
  B.~Alles, G.~Cossu, M.~D'Elia, A.~Di Giacomo and C.~Pica,
  PoS LAT {\bf 2007}, 177 (2007).

\bibitem{Kaneko:2013jla} 
  T.~Kaneko {\it et al.} [JLQCD Collaboration],
  PoS LATTICE {\bf 2013}, 125 (2014)
  [arXiv:1311.6941 [hep-lat]].

\bibitem{Noaki:2014ura} 
  J.~Noaki {\it et al.}  [JLQCD Collaboration],
  PoS LATTICE {\bf 2013}, 263 (2014).

\bibitem{Cossu:2013ola} 
  G.~Cossu, J.~Noaki, S.~Hashimoto, T.~Kaneko, H.~Fukaya, P.~A.~Boyle and J.~Doi,
  arXiv:1311.0084 [hep-lat];\\ 
URL :\url{http://suchix.kek.jp/guido_cossu/documents/DoxyGen/html/index.html}


\bibitem{Borsanyi:2012zs} 
  S.~Borsanyi {\it et al.},
  JHEP {\bf 1209}, 010 (2012)
  [arXiv:1203.4469 [hep-lat]].

\bibitem{Hashimoto:2014gta} 
  S.~Hashimoto, S.~Aoki, G.~Cossu, H.~Fukaya, T.~Kaneko, J.~Noaki and P.~A.~Boyle,
  PoS LATTICE {\bf 2013}, 431 (2014).

\bibitem{Schaefer:2010hu} 
  S.~Schaefer {\it et al.} [ALPHA Collaboration],
  Nucl.\ Phys.\ B {\bf 845}, 93 (2011)
  [arXiv:1009.5228 [hep-lat]].

\bibitem{Bali:2014pva} 
  G.~S.~Bali, S.~Collins, S.~Dürr and I.~Kanamori,
  Phys.\ Rev.\ D {\bf 91}, no. 1, 014503 (2015)
  [arXiv:1406.5449 [hep-lat]].

\bibitem{Aoki:2007ka} 
  S.~Aoki, H.~Fukaya, S.~Hashimoto and T.~Onogi,
  Phys.\ Rev.\ D {\bf 76}, 054508 (2007)
  [arXiv:0707.0396 [hep-lat]].



\bibitem{Brendan}
JLQCD collaboration, in preparation.








\bibitem{Beringer:1900zz} 
  K.~A.~Olive {\it et al.} [Particle Data Group],
  Chin.\ Phys.\ C {\bf 38}, 090001 (2014).


\end{thebibliography}
\end{document}